# Microtesla MRI with dynamic nuclear polarization

Vadim S. Zotev[*], Tuba Owens, Andrei N. Matlashov,
Igor M. Savukov, John J. Gomez, Michelle A. Espy

Los Alamos National Laboratory, Applied Modern Physics Group, MS D454, Los Alamos, NM 87545, USA

**Abstract**

Magnetic resonance imaging at microtesla fields is a promising imaging method that combines the pre-polarization technique and broadband signal reception by superconducting quantum interference device (SQUID) sensors to enable in vivo MRI at microtesla-range magnetic fields similar in strength to the Earth magnetic field. Despite significant advances in recent years, the potential of microtesla MRI for biomedical imaging is limited by its insufficient signal-to-noise ratio due to a relatively low sample polarization. Dynamic nuclear polarization (DNP) is a widely used approach that allows polarization enhancement by two-four orders of magnitude without an increase in the polarizing field strength. In this work, the first implementation of microtesla MRI with Overhauser DNP and SQUID signal detection is described. The first measurements of carbon-13 NMR spectra at microtesla fields are also reported. The experiments were performed at the measurement field of 96 microtesla, corresponding to Larmor frequency of 4 kHz for protons and 1 kHz for carbon-13. The Overhauser DNP was carried out at 3.5 – 5.7 mT field using rf irradiation at 120 MHz. Objects for imaging included water phantoms and a cactus plant. Aqueous solutions of metabolically relevant sodium bicarbonate, pyruvate, alanine, and lactate, labeled with carbon-13, were used for NMR studies. All the samples were doped with TEMPO free radicals. The Overhauser DNP enabled nuclear polarization enhancement by factor as high as −95 for protons and as high as −200 for carbon-13, corresponding to thermal polarizations at 0.33 T and 1.1 T fields, respectively. These results demonstrate that SQUID-based microtesla MRI can be naturally combined with Overhauser DNP in one system, and that its signal-to-noise performance is greatly improved in this case. They also suggest that microtesla MRI can become an efficient tool for in vivo imaging of hyperpolarized carbon-13, produced by the low-temperature dissolution DNP.

## 1. Introduction

Magnetic resonance imaging is characterized by inherently low sensitivity, resulting from the fact that the energy of a nuclear spin in a typical MRI magnetic field is considerably lower than its thermal energy at room temperature. For a system of $I$=1/2 nuclear spins with gyromagnetic ratio $\gamma_I$, the thermal equilibrium polarization, i.e. the difference between populations of the two Zeeman energy levels divided by the sum of their populations, in magnetic field $B_0$ at temperature $T$ is $P_{th} \approx \gamma_I \hbar B_0/(2kT)$. This value is very low under conventional MRI conditions. For example, for proton spins at 3 T field and room temperature, $P_{th} \sim 1\cdot10^{-5}$. The sample magnetization, $M_0 \approx N\gamma_I^2\hbar^2 B_0/(4kT)$, where $N$ is the number of spins per unit volume, scales linearly with $B_0$ and quadratically with $\gamma_I$. Because the signal, induced in a Faraday receiver coil, increases linearly with Larmor frequency $\omega_0=\gamma_I B_0$, the signal-to-noise ratio (SNR) in conventional NMR/MRI is proportional to $\gamma_I^3 B_0^2$, if the noise is frequency independent. If the noise is dominated by the receiver's thermal noise, which increases with frequency because of the skin effect, the SNR scales as $\gamma_I^{11/4} B_0^{7/4}$ at high fields [1]. The approximately cubic dependence of the SNR on $\gamma_I$ means that the sensitivity of conventional NMR/MRI to $^{13}$C spins is 64 times lower than to proton spins ($\gamma_C/\gamma_H$=0.251) at the same magnetic field, temperature, and spin concentration.

Various physical approaches have been proposed to improve SNR of NMR/MRI in liquids and biological tissues without increasing the strength of the magnetic field $B_0$. They typically employ a combination of two different magnetic fields. In the well known pre-polarization technique [2], the sample is pre-polarized by a magnetic field $B_p$, and NMR signal is measured at a lower measurement field $B_m$ after the pre-polarizing field is rapidly removed. The SNR in this technique is proportional to $\gamma_I^3 B_p B_m$ for Faraday detection, and is improved compared to conventional NMR at the low field $B_m$. This method has been traditionally used to perform NMR/MRI in the Earth magnetic field of ~50 μT (e.g. [3−5]).

Recently, a new approach to enhance SNR of magnetic resonance at low fields was introduced [6, 7]. It is referred to as SQUID-based microtesla NMR/MRI or ultralow-field (ULF) NMR/MRI. This approach combines the pre-polarization technique and broadband signal reception

---

using SQUID sensors [8, 9] with untuned input circuits that act as frequency independent flux-to-voltage transducers. Modern low-$T_c$ SQUID systems provide frequency independent noise performance above ~1 Hz with magnetic field noise levels as low as ~1 fT/Hz$^{1/2}$. The sensitivity of SQUID signal detection in ULF NMR/MRI does not depend on Larmor frequency in the measurement field $B_m$, so the SNR is proportional to $\gamma_I^2 B_p$, and is enhanced substantially compared to the approach with pre-polarization and Faraday detection at low fields. Moreover, because the NMR line broadening due to magnetic field inhomogeneity scales linearly with the field strength (for a fixed relative inhomogeneity), narrow NMR lines with high SNR [6] can be obtained at a microtesla-range $B_m$ field generated by a simple and inexpensive coil system. The NMR sensitivity for $^{13}$C in this method is 16 times lower than for $^1$H under the same conditions. Significant progress was made in recent years in the development and applications of both ULF NMR [10−17] and ULF MRI [18−25]. However, the potential of ULF MRI for medical imaging is limited by the efficiency of pre-polarization. For example, polarization of proton spin populations at the pre-polarizing field of 30 mT, used for human brain imaging in [21], is ~1·10$^{-7}$, i.e. 100 times lower than at 3 T field of conventional MRI. It becomes increasingly difficult to generate pulsed $B_p$ fields above ~0.1 T within large sample volumes using resistive coils, because of the higher energy consumption and heat dissipation.

Dynamic nuclear polarization (DNP), based on the well known Overhauser effect [26], is an efficient way to substantially enhance nuclear polarization without increasing the strength of the polarizing field. The DNP procedure usually involves addition of unpaired electron spins (e.g. free radicals) to the studied sample and rf (microwave) irradiation of the resulting two-spin system at the frequency of the electron spin resonance (ESR). Saturation of the ESR resonance leads to a transfer of polarization from electron to nuclear spins, and creates a non-equilibrium nuclear polarization, that can be much larger than the thermal equilibrium polarization at the same magnetic field [27, 28]. The DNP enhancement factor is proportional to $(\gamma_S/\gamma_I) \cdot F(B_p)$, where $\gamma_S$ is the gyromagnetic ratio for electron, and $F(B_p)$ describes dependence of the enhancement on the magnetic field strength, the spin system properties, and the degree of saturation of the ESR resonance. If the NMR/MRI procedure includes DNP enhancement at the field $B_p$ and signal measurement at the field $B_m$, the SNR scales as $\gamma_S \gamma_I^2 B_p B_m$ for Faraday detection, and as $\gamma_S \gamma_I B_p$ for SQUID detection. These estimates are based on the assumption that $F(B_p)$ is a slow varying function of $B_p$, at least for a field range of interest, and noise is frequency independent. Because the electron gyromagnetic ratio $\gamma_S$ is much larger by absolute value than $\gamma_I$ for protons ($|\gamma_S/\gamma_H|=658$), DNP allows enhancement of $^1$H polarization by two orders of magnitude. The sensitivity of DNP-enhanced NMR/MRI for $^{13}$C is improved further, and can be as high as 1/4 of its sensitivity for $^1$H (under the same conditions) when SQUID detection is used. Moreover, the broadband SQUID reception allows simultaneous measurement of $^1$H and $^{13}$C NMR signals with the same sensor.

Overhauser-enhanced MRI (e.g. [29−33]), also referred to as proton-electron double resonance imaging (PEDRI), is an imaging technique based on Overhauser DNP in liquids, which combines the advantages of ESR and proton MRI. DNP in this method is typically performed at a magnetic field of several mT using rf irradiation frequency of the order of 100 MHz. The specific absorption rate (SAR) is thus sufficiently low to allow in vivo DNP. One application of Overhauser-enhanced MRI is in vivo imaging of free radicals, either endogenous or injected [30]. Another important application is dynamic in vivo oxymetry, which can be used for detection and monitoring of oxygen deficiency in tumors [32].

Recently, a new and very promising implementation of DNP, that allows important imaging applications, was proposed [34]. It is called DNP hyperpolarization or dissolution DNP. In this approach, DNP is performed for a sample in solid state at a low temperature (~1 K) and strong magnetic field (~3 T) using microwave irradiation frequency of the order of 100 GHz [34]. Because the thermal equilibrium polarization $P_{th}$ is inversely proportional to $kT$, the reduction in temperature from ~300 K to ~1 K increases the thermal polarization by two orders of magnitude. Combined with the DNP polarization enhancement by another two orders of magnitude, this leads to an overall polarization increase by more than 10000 times over the thermal polarization level at 3 T field and room temperature [34]. The predominant DNP mechanism in this case is thermal mixing [35]. Rapid dissolution of the sample preserves the nuclear polarization in liquid phase, and produces a hyperpolarized liquid that can be separated from the free radicals and injected in the bloodstream. Applicability of this technique is limited to nuclear spins with long relaxation time $T_1$ in liquid state (>20 s), such as $^{13}$C and $^{15}$N at certain positions within organic molecules. $^{13}$C polarization levels as high as 40% have been demonstrated after the DNP hyperpolarization and sample dissolution [34]. The hyperpolarization technique has been successfully used for angiography, perfusion mapping, and interventional MRI with unprecedented SNR levels [36]. It has also enabled real-time metabolic imaging [37], in which relative concentrations of metabolites of an injected hyperpolarized endogenous substance, such as [1-$^{13}$C] pyruvate, are monitored by means of MRI spectroscopy of $^{13}$C. Such imaging has shown promise for cancer research, where it can be used for noninvasive detection and grading of tumors, and for early assessment of tumors' response to treatment (e.g. [38]).

It has been suggested that DNP mechanisms "promise to form a perfect complement to microtesla MRI" [39]. In this work, we demonstrate microtesla MRI with Overhauser DNP enhancement and SQUID signal detection. We also report Overhauser-enhanced NMR of $^{13}$C in metabolically relevant substances at microtesla magnetic fields. Finally, we present a detailed argument for integration of the DNP hyperpolarization technique with SQUID-based microtesla MRI.

## 2. Methods

### 2.1 Overhauser enhancement

It is instructive to begin the discussion of Overhauser DNP by considering the standard four-state diagram [27, 28] that shows energy levels for a system of two 1/2 spins in a strong magnetic field (Fig. 1). Here, $S$ is the unpaired electron spin ($\gamma_S < 0$), and $I$ is a nuclear spin. The diagram in Fig. 1 also includes spin-flip processes and the corresponding transition probabilities. Irradiation of this system at the ESR frequency equilizes electron spin populations of levels 1 and 3, as well as 2 and 4, and saturates electron spin-flip transitions 1↔3 and 2↔4. This leads to a relaxation via transitions 1↔4 and 2↔3, that involve flipping of both spins, provided that either $w_0$ or $w_2$ transition probability is sufficiently high. Such cross-relaxation enables transfer of polarization from electron to nuclear spins.

According to the theory of Overhauser effect, the expectation values $\langle I_z \rangle$ and $\langle S_z \rangle$ for the spin system in Fig. 1 are related as follows [27, 28]:

$$\frac{\langle I_z \rangle - I_0}{I_0} = -\rho f \left( \frac{\langle S_z \rangle - S_0}{I_0} \right) \quad (1)$$

Here, $I_0$ and $S_0$ are expectation values of $I_z$ and $S_z$ in thermal equilibrium, $\rho$ is the coupling factor, and $f$ is the leakage factor. These two factors are defined by the following expressions:

$$\rho = \frac{w_2 - w_0}{w_0 + 2w_1 + w_2} \quad (2)$$

$$f = \frac{w_0 + 2w_1 + w_2}{w_0 + 2w_1 + w_2 + w_1^0} = 1 - \frac{T_1}{T_{10}} \quad (3)$$

The coupling factor $\rho$ ranges from −1 for pure scalar coupling ($w_0 > w_2$) between the electron and the nuclear spins to +0.5 for pure dipolar coupling ($w_2 > w_0$). The leakage factor $f$ takes values between 0 and 1, and depends on the ratio of the longitudinal relaxation times for nuclear spins in the presence ($T_1$) and in the absence ($T_{10}$) of unpaired electron spins.

To describe the degree of saturation of the ESR resonance upon rf irradiation, the saturation factor $s$ is commonly introduced:

$$s = \frac{S_0 - \langle S_z \rangle}{S_0} \quad (4)$$

The saturation factor changes between 0 for thermal equilibrium and 1 for complete saturation of all ESR transitions, i.e. $\langle S_z \rangle = 0$. It depends on the power of rf irradiation, the relaxation properties of the electron spins, and the structure of the ESR spectrum at a given magnetic field strength.

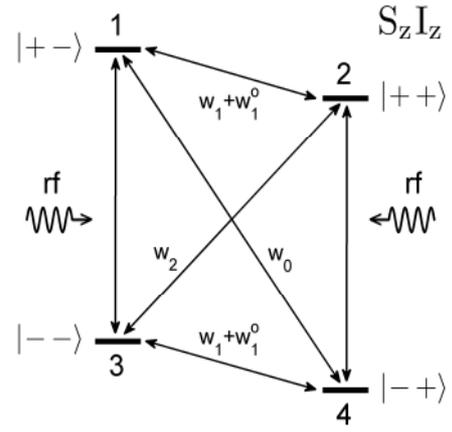

**Fig. 1.** Diagram of energy levels and transitions in a system of two spins ($S=½$, $I=½$) in a strong magnetic field.

The enhancement of nuclear spin polarization due to Overhauser effect is typically characterized by the enhancement factor $E$, defined as $E = \langle I_z \rangle / I_0$. It can be expressed as follows:

$$E = 1 - \rho f s \frac{|\gamma_S|}{\gamma_I} \quad (5)$$

This expression is obtained from Eq. (1) using Eq. (4) and an assumption that both $I_0$ and $S_0$ correspond to the same magnetic field $B_0$. This assumption is often violated in real systems at low magnetic fields. When it holds, Eq. (5), predicts the maximum polarization enhancement for protons to be 659 in the case of pure scalar coupling and −328 for pure dipolar coupling. The negative $E$ means that the non-equilibrium magnetization vector, created by Overhauser DNP, has the direction opposite to that of the thermal equilibrium magnetization in the external magnetic field $B_0$.

Because only one ESR transition can, in practice, be effectively saturated at a time, the maximum achievable enhancement is lower, if the ESR spectrum has several lines. Assuming that the ESR lines are Lorentzian and saturation of one transition does not affect the others, one can use the following approximate expression for the saturation factor [40]:

$$s = \frac{1}{n}\left(\frac{\alpha P_{rf}}{1 + \alpha P_{rf}}\right) \quad (6)$$

Here, $n$ is the number of lines in the ESR spectrum, $P_{rf}$ is the rf irradiation power, and $\alpha$ is a constant related to relaxation properties of electron spins and the rf resonator characteristics. For $n=3$, a complete saturation of one ESR resonance, $\langle S_z \rangle = (2/3)S_0$, leads to the saturation factor value $s=1/3$ and provides the maximum enhancement of −109 in the case of pure dipolar coupling.

*2.2 Overhauser DNP at low fields*

Nitroxide free radicals [41, 42] have been widely used as sources of unpaired electron spins for ESR and DNP, and their Overhauser DNP properties have been investigated in detail (e.g. [40, 43−45]). The readers are referred to [43] for a rigorous analytical treatment, and to [44] for a thorough physical discussion of the low-field case.

In nitroxide radicals, such as TEMPO ($C_9H_{18}NO$), the unpaired electron spin $S$ is coupled to the nuclear spin of the nitrogen atom ($K$=1 for $^{14}N$), and the energy spectrum is split into six levels. As the external magnetic field $B_0$ is reduced below several mT, the hyperfine term $AS·K$ becomes comparable to and eventually dominates the Zeeman term $-\gamma_S S_z B_0$ in the Hamiltonian. The unpaired electron spin experiences a strong local magnetic field from the nitrogen spin, and the energies of the six levels are no longer proportional to $B_0$. The equilibrium expectation value $S_0$ in Eq. (1) is no longer the expectation value in the field $B_0$, and Eq. (5) for Overhauser enhancement does not hold. Because the expectation value $I_0$ for the nuclear spins in solution scales linearly with $B_0$, the enhancement factor $E$ increases drastically as the magnetic field is lowered, according to Eq. (1). It was predicted to be as high as ±2000 in the Earth magnetic field [43]. This effect has limited practical significance, however, because the *absolute* polarization levels, achieved with Overhauser DNP at microtesla-range magnetic fields, are nevertheless quite low.

The electron and nitrogen spins in TEMPO are strongly coupled at low magnetic fields, and this two-spin system has 10 allowed ESR transitions [43, 44]. Eight of them are π transitions that can be induced by rf magnetic fields perpendicular to the static magnetic field $B_0$. The other two are σ transitions that are induced by rf irradiation fields parallel to $B_0$. However, only three transitions ($T_{16\pi}$, $T_{25\pi}$, and $T_{34\pi}$) out of ten have significant transition probabilities at magnetic fields above ~2 mT. Thus, the ESR spectrum of TEMPO free radicals in the field range 2 – 10 mT, most commonly used for Overhauser DNP, has three main resonance lines. The maximum polarization enhancement $E_{max}$ for solvent water protons, obtained via a complete saturation of any one of these resonances, is predicted to be of the order of −100 [43].

*2.3 Instrumentation*

Our instrumentation for 3D ULF MRI has been described in detail before [19−22]. We have used this system to acquire images of the human brain [21] and study parallel imaging at microtesla fields [20]. The system includes seven SQUID channels in a liquid He cryostat. They are characterized by magnetic field noise as low as 1.2 fT/Hz$^{1/2}$ at 1 kHz inside a magnetically shielded room (MSR). The modified system for ULF MRI with DNP, employed in the present work, is depicted schematically in Fig. 2. Its main difference from the previously used set-up is that the relatively heavy pre-polarization coils, cooled with liquid nitrogen [19, 22], are replaced with a pair of light Helmholtz coils and a thin rf antenna.

The new $B_p$ coils are designed to generate a magnetic field of several mT. They are square with the side length of 61 cm, 2.5 cm × 2.5 cm cross section, and 33.5 cm spacing (Fig. 2A). The relative non-uniformity of the $B_p$ field is less than 0.1% within 10 cm wide sample space, and the field strength is 0.57 mT at 1 A. Because the magnetic field stability is essential for DNP experiments, the $B_p$ coils are driven by dc power supplies (Sorensen DLM 40-15) in current control mode. The $B_p$ current is turned on and off with specially designed solid state switches [19]. An additional mechanical relay is used in the present set-up to completely disconnect the $B_p$ coils from the power supplies before each measurement (after the pre-polarization), and thus prevent the power line noise from entering the system. The other four sets of coils in Fig. 2A are used to generate the μT-range measurement field $B_m$ and three gradients, $G_x$=d$B_z$/d$x$, $G_y$=d$B_z$/d$y$, and $G_z$=d$B_z$/d$z$, for 3D Fourier imaging as described previously [19].

The rf antenna is used for irradiation of the sample at the ESR frequency of 120 MHz. Its schematic is shown in Fig. 2B. We chose the antenna design with a surface coil and a parallel transmission line [33, 46], that is widely used in ESR and Overhauser-enhanced MRI. The resonant circuit in this design is formed by the inductance of the surface coil and the distributed capacitance of the parallel line, which consists of two coaxial cables of length $l$. The parallel line has an open-circuit termination and is balanced by the λ/2 balun, formed by two coaxial pieces of λ/4 length (Fig. 2B). The balun ensures that voltages at the ends of the parallel line are opposite and symmetric with respect to the ground. The $C_T$ capacitor is used for frequency tuning, and $C_M$ – for impedance matching.

Because larger surface coils allow irradiation of larger objects, we built three rf antennas with coil diameters of 20 mm, 40 mm, and 60 mm, using the standard RG 223/U coaxial cable. Parameters of the antennas were calculated according to formulas in [46]. The parallel transmission line length was computed to be $l$=320 mm, 199 mm, and 95 mm at 120 MHz for the 20 mm, 40 mm, and 60 mm antennas, respectively. The balun included two cable pieces of λ/4=412 mm length each. To reduce heating, the antenna coils were made of 2 mm thick copper wire, and the outer insulation layer was removed from the coaxial cables. Variable

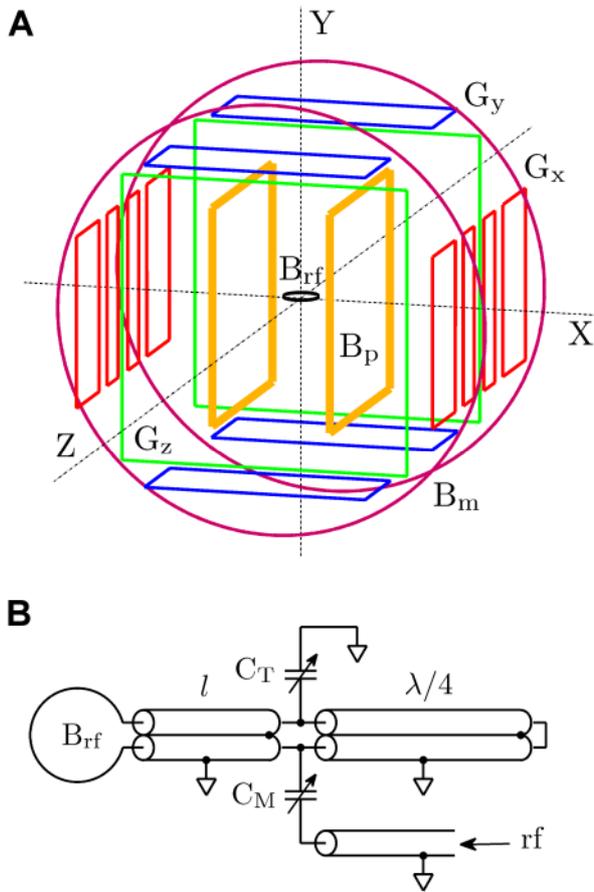

**Fig. 2.** A) Schematic of the coil system for 3D ULF MRI with DNP. B) Schematic of the antenna used for rf irradiation of the sample.

0.8…10 pF rf-rated air capacitors were used for $C_T$ and $C_M$.

To measure generation efficiency of the antennas, we employed a small (3 mm diameter) probe coil in series with a small-size 50 ohm resistor, placed in front of the antenna and connected via a $\lambda/2$ length cable to a high-frequency spectrum analyzer (HP 8566B). The power of the rf signal induced in the probe coil is related to the amplitude of the $B_{rf}$ field produced by the antenna. The generation efficiency of the unloaded antenna coils was measured to be 28 $\mu T/W^{1/2}$, 14 $\mu T/W^{1/2}$, and 8.4 $\mu T/W^{1/2}$ at 120 MHz for the 20 mm, 40 mm, and 60 mm antennas, respectively. The measured values of the quality factor $Q$ for the three antennas were, respectively, 90, 101 and 78. The 120 MHz signal for rf irradiation was provided by a general-purpose signal generator (HP 8662A) and amplified by a high-power rf amplifier (Amplifier Research 1000LP). A directional coupler (−30 dB) was installed at the output of the amplifier to monitor the applied rf power with the spectrum analyzer. High-power rf relays were used to turn the signal on and off, with one relay included between the rf generator and the amplifier, and the other – between the amplifier and the antenna. To reduce interference penetration, the latter relay was placed in the feed-through opening of the MSR, and its rf ground was electrically connected to the MSR wall.

To optimize DNP performance of the system, it is essential to maximize the rf power supplied to the sample by properly tuning the antenna. The tuning was performed in three steps before each experiment. First, the antenna was positioned outside the ULF MRI system, and the small probe coil, connected to the spectrum analyzer, was placed in front of it. The $C_T$ and $C_M$ capacitors were adjusted to tune the resonance frequency and maximize the rf signal induced in the probe coil. Second, the antenna was installed inside the ULF MRI system, and further adjustments were made while the probe coil signal was observed. Third, the sample was placed inside (or around) the antenna, and the $C_M$ capacitor was adjusted again to maximize the DNP-enhanced ULF NMR signal, measured by the SQUID instrumentation and monitored with a general-purpose spectrum analyzer (Stanford Research Systems SR760). This last step requires that the $B_p$ field strength be set to a value close to a DNP resonance.

*2.4 Experimental parameters*

The experimental protocol, used in the present work for Overhauser-enhanced ULF MRI, is illustrated in Fig. 3A, and the protocol for ULF NMR of $^{13}$C is shown in Fig. 3B. In both protocols, each measurement step begins with the pre-polarization stage, during which the sample is subjected to rf irradiation in the presence of the constant pre-polarizing field $B_p$. The typical rf power of 100 W was supplied to the antenna during the irradiation. The pre-polarization field strength was $B_p$=3.52 mT in $^1$H imaging experiments, and $B_p$=5.75 mT in $^{13}$C NMR measurements. The switch-off time for the $B_p$ field was 2 ms at 10 A current. The measurement field $B_m$ in all the experiments was set to 96 $\mu$T, corresponding to the Larmor precession frequency $f_H$=4096 Hz for $^1$H and $f_C$=1030 Hz for $^{13}$C.

The antenna coil was positioned horizontally (XZ plane) during the imaging experiments, and the object was placed inside the coil as shown in Fig. 4 (left). The measurement field $B_m$ was applied after the removal of the $B_p$ field perpendicular to the polarization vector, inducing spin precession [19]. The echo signal was formed by simultaneous reversal of the $B_m$ field and the frequency encoding gradient $G_x$ (Fig. 3A). The pre-polarization time $t_p$ used for $^1$H imaging was either 1.0 s or 1.5 s. The gradient encoding and signal acquisition times were $t_g$=0.25 s and $t_a$=0.5 s, respectively. The frequency encoding gradient $G_x$ changed between ±47 $\mu$T/m (±20 Hz/cm for $^1$H). The phase encoding gradient $G_z$ had $N_z$=65 different values, that were equally spaced and symmetric with respect to $G_z$=0, with the maximum value $G_{z,max} = -G_{z,min}$= 47 $\mu$T/m (20 Hz/cm). These imaging parameters provided 1 mm × 1 mm resolution in the horizontal plane. In 3D imaging experiments, $N_y$=11 phase encoding steps were taken with the maximum gradient value $G_{y,max} = -G_{y,min}$ = 9.4 µT/m (4.0 Hz/cm for $^1$H), providing 5 mm resolution in the vertical (Y) direction. The measured signal was digitized at 21.3 kS/s sampling rate. All ULF MR images with Overhauser enhancement were acquired in a single scan, i.e. without any signal averaging. Because the samples were relatively small (Fig. 4), the data from only one SQUID channel (Ch 1) out of seven were used in the analysis.

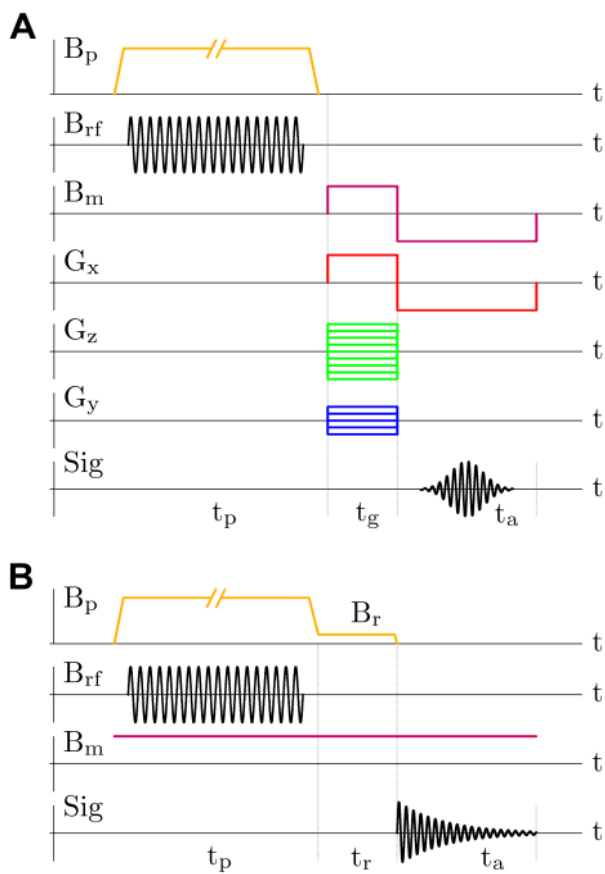

**Fig. 3.** A) Experimental protocol for ULF MRI with Overhauser DNP. B) Protocol for ULF NMR with DNP.

For the $^{13}$C NMR measurements, the antenna coil was positioned vertically in the XY plane. The sample consisted of four vials, each containing 2 ml of solution, placed on both sides of the coil as shown in Fig. 4 (right). A forced air flow around the sample was used for cooling. The measurement field $B_m$ was kept constant throughout the experiment (Fig. 3B), and spin precession was induced by rapidly switching off the $B_p$ field (or the $B_r$ field if $t_r$>0). This approach was chosen for the NMR measurements, because the rapid application of the $B_m$ field in the ULF MRI procedure (Fig. 3A) leads to a magnetic relaxation of the MSR in the direction of $B_m$, which remains sufficiently strong for about 100 ms to cause a visible drift in Larmor frequency. This drift is not important for imaging, because of the relatively strong $G_x$ gradient (Fig. 3A), but it affects the shape of the measured NMR line. The $B_p$ switching off in 2 ms was non-adiabatic for $^{13}$C spins ($f_C$~1 kHz), because virtually no $^{13}$C signal loss was observed in the case of steady $B_m$ field compared to the protocol with pulsed $B_m$. The signal from proton spins ($f_H$~4 kHz), however, decreased by a factor of 2 at the steady $B_m$, and this effect was taken into account when comparing the NMR signal strengths for $^1$H and $^{13}$C. The relaxation field $B_r$, generated by the same $B_p$ coils during a variable delay time $t_r$, was used to measure the longitudinal relaxation time $T_1$ at the $B_r$ field as in [22]. The pre-polarization time $t_p$ was 5 s for $^{13}$C sodium bicarbonate, and 10 s for the other $^{13}$C-labeled chemicals. Each NMR measurement was repeated 50-100 times, and the resulting spectra were averaged.

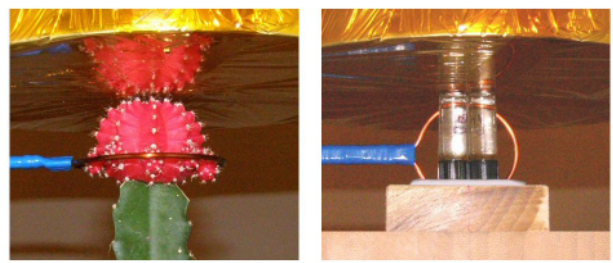

**Fig. 4.** *Left*: position of the rf antenna for imaging experiments (the 60 mm antenna is shown). *Right*: position of the rf antenna for NMR measurements of carbon-13 (the 40 mm antenna is shown).

*2.5 Materials*

The Overhauser DNP experiments, described in this work, were performed using TEMPO (2,2,6,6-tetramethylpiperidine 1-oxyl) free radicals, purchased from MP Biomedicals (www.mpbio.com). We also used $^{13}$C labeled substances, including $^{13}$C sodium bicarbonate $NaHCO_3$, [1-$^{13}$C] sodium pyruvate $CH_3COCOONa$, [1-$^{13}$C] L-alanine $CH_3CH(NH_2)COOH$, and [1-$^{13}$C] sodium L-lactate $CH_3CH(OH)COONa$. The pyruvate, alanine, and lactate were labeled with $^{13}$C in the carboxyl −COO− position (also referred to as $C_1$) with isotopic enrichment of 99%. These materials were purchased from Cambridge Isotope Laboratories, Inc. (www.isotope.com). As mentioned in Sec. 1, the hyperpolarized [1-$^{13}$C] pyruvate is used in metabolic imaging studies, where its conversion to alanine and lactate is monitored as a function of time [37]. The $T_1$ relaxation time for the carboxyl $^{13}$C in pyruvate is longer than 20 s. All the chemicals were dissolved in de-ionized water (without degassing), and a fresh solution was prepared before each experiment.

The $^3J_{CH}$ and $^2J_{CH}$ coupling constants for the carboxyl $^{13}$C in [1-$^{13}$C] pyruvate, alanine, and lactate are in the range of 3-5 Hz. Because these values are three orders of magnitude lower than the difference in Larmor frequencies between $^1$H and $^{13}$C (~3 kHz in our experiments), these systems are weakly coupled, and analysis of their $^{13}$C spectra is straightforward. In pyruvate, interaction of $^{13}$C with three protons in the methyl CH$_3-$ group (also referred to as C$_3$ or C$_\beta$ group) leads to a quartet of lines with $^3J_{CH}$~1.3 Hz [47]. In alanine, $^{13}$C is coupled to three methyl protons with $^3J_{CH}$~4.2 Hz and to a proton in the methine $-$CH$-$ group (C$_2$ of C$_\alpha$) with $^2J_{CH}$~5.0 Hz [47]. Thus, the $^{13}$C spectrum of alanine is an octet. For lactate, the corresponding coupling constants are $^3J_{CH}$~4.2 Hz and $^2J_{CH}$~3.5 Hz [47], so the spectrum is also an octet with a slightly smaller splitting. Because three pairs of lines at the center of the octet spectrum have line separations of only ~0.8 Hz for alanine and ~0.7 Hz for lactate, the $^{13}$C spectra of these two systems may appear as quintets when the spectral resolution is not very high. In [48], the *J*-coupling spectra of $^{13}$C in [1-$^{13}$C] alanine and lactate, measured at 3 T with 1 Hz resolution, look like quintets with *J*~4.5 Hz and *J*~3.75 Hz line splittings, respectively. These data suggest that alanine and lactate can be distinguished based on their *J*-coupling spectra. A much higher spectral resolution at microtesla-range magnetic fields [4] should allow highly accurate measurements of $^{13}$C *J*-coupling spectra for these materials.

## 3. Results

### 3.1 Overhauser DNP performance

Fig. 5A exhibits the resonance structure of Overhauser DNP probed as a function of the pre-polarizing field $B_p$. It was measured for 2 mM (i.e. 2 mM/liter) water solution of TEMPO free radicals at a fixed rf irradiation frequency of 120 MHz using the 40 mm antenna at 20 W input power. The three peaks at 2.15 mT, 3.52 mT, and 5.77 mT fields result from the hyperfine splitting of the TEMPO radicals' ESR resonance, and correspond to the T$_{16\pi}$, T$_{25\pi}$, and T$_{34\pi}$ transitions (Sec. 2.2). We performed similar measurements for 1M water solution of $^{13}$C sodium bicarbonate, doped with 16 mM of TEMPO. The $^{13}$C resonances were broader, but their peak positions were very close to those for $^1$H, with the tallest peak at $B_p$=5.75 mT.

Fig. 5B compares the DNP enhancement factors for the three antennas of 20 mm, 40 mm, and 60 mm diameter. Each antenna was placed horizontally 25 mm below the cryostat. A small vial (9 mm diameter, 2 ml volume) with 2 mM TEMPO solution was put at the center of the antenna, and its position with respect to the cryostat was always the same. The enhancement factor *E* was determined as the ratio of the ULF NMR signal intensities measured at $B_p$=3.52 mT with and without rf irradiation. The pre-polarization time in these experiments was $t_p$=1.5 s, and the $T_1$ relaxation time for 2 mM TEMPO solution was measured to be 0.79±0.05 s at 3.52 mT. According to Fig. 5B, the polarization enhancement depends strongly on the antenna coil size at low power levels, with the 20 mm antenna providing *E* as large as −29 at 5W. As the input power increases, however, the saturation regime is gradually approached, and the differences in *E* among the three antennas become less pronounced. Therefore, both the 40 mm and the 60 mm antennas can be efficiently used for ULF MRI with DNP, provided that the applied rf power is sufficiently high. The largest polarization enhancement, observed for $^1$H at $B_p$=3.52 mT using the 20 mm antenna, was *E*= −95, which corresponds to the equivalent polarization field of 0.33 T. The DNP experiments, reported below in Sec. 3.2 and Sec. 3.3, were carried out at either 3.52 mT or 5.75 mT field rather than at 2.15 mT to facilitate real-time monitoring of the unenhanced $^1$H signal and achieve more accurate tuning of the system as described in Sec. 2.3 above.

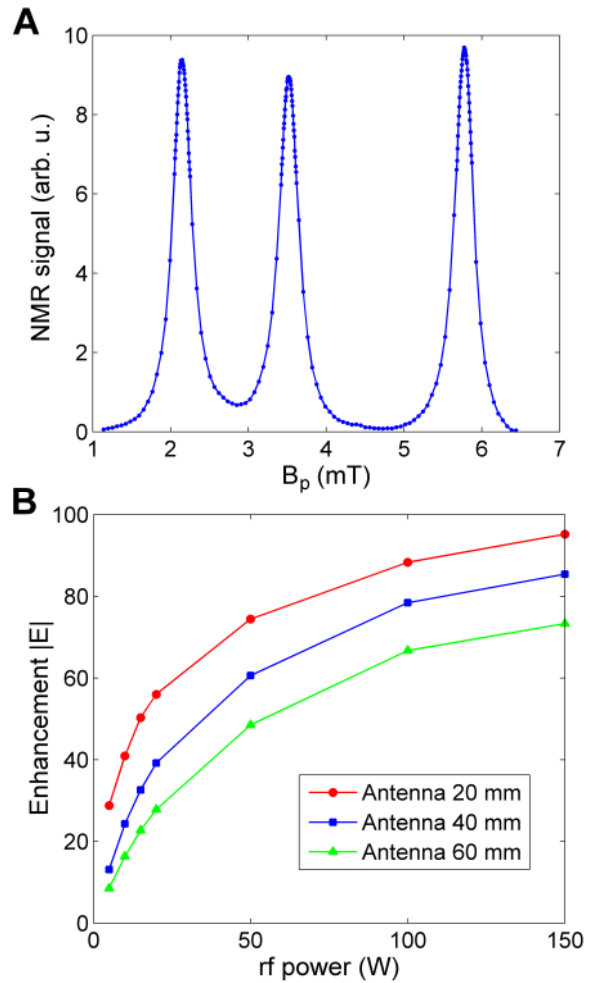

**Fig. 5.** A) DNP resonance structure for 2 mM water solution of TEMPO free radicals. B) Absolute value of the polarization enhancement factor *E* measured as a function of applied rf power.

The plots of $1/(1-E)$ vs $1/P_{rf}$ using the data in Fig. 5B yielded straight lines with the same intercept $1/(1-E_{max})$, corresponding to $E_{max}= -101$. The leakage factor was $f=0.8$ according to Eq. (3). Using Eqs. (5) and (6), we then determined the ratio $(\rho/n)=0.194$. For $n=3$, this gives the coupling factor value $\rho=0.58$, which is higher than the theoretical value $\rho=0.5$ for pure dipolar coupling. Such discrepancy has been observed before (e.g. [40]). It reflects the fact that Eqs. (5) and (6) are based on the four-level model in Fig. 1, which is not entirely accurate for nitroxide radicals even at high magnetic fields, as discussed in [45].

The strong rf pulses, employed in the present work, did not have any negative effect on the performance of the SQUID instrumentation. This is because the cryostat in our ULF MRI system is screened with an rf shield made of gold-plated mylar (Fig. 4), and special cryoswitches [19] are used to disconnect the pick-up coils from the SQUID sensors during the pre-polarization. However, the λ/2 balun part of the antenna (Fig. 2B) provides a return path for Johnson currents, which causes an increase in thermal noise level. For example, when the unloaded antenna coil was placed horizontally 25 mm below the bottom of the cryostat (Fig. 4, left), the magnetic field noise spectral densities were measured to be 1.25 fT/Hz$^{1/2}$, 1.52 fT/Hz$^{1/2}$, and 1.88 fT/Hz$^{1/2}$, for the 20 mm, 40 mm, and 60 mm diameter antennas, respectively. When the antenna was moved up to 18 mm below the cryostat, these noise values increased, respectively, to 1.34 fT/Hz$^{1/2}$, 1.82 fT/Hz$^{1/2}$, and 2.36 fT/Hz$^{1/2}$. In the absense of an rf antenna, the noise spectral density for Ch 1 was 1.2 fT/Hz$^{1/2}$ at 1 kHz. Thus, the actual SNR improvement in ULF NMR/MRI with Overhauser DNP is somewhat lower than the polarization enhancement $E$. This effect can be mitigated by positioning the rf antenna coil vertically (Fig. 4, right). For the 40 mm antenna, the noise level was measured to be 1.38 fT/Hz$^{1/2}$ in this case. Such configuration was successfully used for $^{13}$C NMR measurements (Sec. 3.3), but it is not always convenient for imaging. Modifications to the antenna design that could reduce its thermal noise without compromising its rf generation efficiency would be beneficial.

*3.2 Overhauser-enhanced ULF MRI*

The first images, acquired by Overhauser-enhanced MRI with SQUID signal detection, are exhibited in Fig. 6. Three vials of 14 mm, 30 mm, and 47 mm internal diameter were filled with 2 mM water solution of TEMPO, and placed inside the 20 mm, 40 mm, and 60 mm antennas, respectively. The imaging experiments were performed as described in Sec. 2.4 with the pre-polarization time $t_p=1.5$ s. According to Fig. 6, the DNP enhancement leads to a dramatic improvement in image quality. The image of the 47 mm phantom shows some signal reduction toward the phantom edges, because sensitivity of the 37 mm diameter SQUID pick-up coil is decreased in those areas [19]. The DNP images in Fig. 6 are characterized by enhancement factors $|E|\sim60\ldots80$ (near the image center), which, at $B_p=3.52$ mT, correspond to equivalent polarization fields of 0.2…0.3 T. These results demonstrate that Overhauser-enhanced ULF MRI, implemented with a pair of light Helmholtz coils and a thin rf antenna, can produce images that would require the use of a heavy and energy-consuming pre-polarization magnet in conventional ULF MRI.

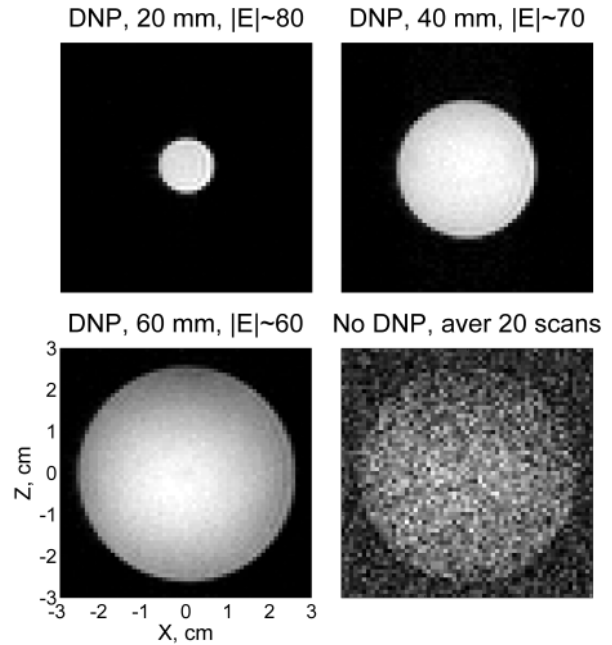

**Fig. 6.** DNP-enhanced 2D images of phantoms containing water solution of TEMPO. The images were acquired at 96 μT field in a single scan with 1 mm × 1 mm resolution. The image without DNP was obtained by averaging 20 scans.

We also used our system for ULF MRI with Overhauser DNP to perform 3D imaging of the insides of a cactus. We chose a cactus plant, because it contains a soft tissue that readily absorbs water. A healthy moon cactus (*Gymnocalycium mihanovichii Hibotan*), that had been regularly watered, was injected with 6 mM water solution of TEMPO, and placed under the bottom of the cryostat inside the 60 mm antenna (Fig. 4, left). The pre-polarization time at each imaging step was $t_p=1.0$ s. Because 11 image layers were acquired with 1 mm × 1 mm in-plane resolution and 5 mm resolution in the vertical dimension, the total imaging time for a single scan was about 20 min. Fig. 7 exhibits the resulting image, which reproduces the cactus shape and shows some details of its internal anatomy.

In order to estimate time dependence of the polarization enhancement factor $E$, we monitored the NMR signal from a similar cactus, also injected with TEMPO. The initial enhancement value was found by extrapolation to be $|E|\sim 24$. The actual enhancement was likely to be higher, because the maximum DNP enhancement occurred at the antenna level 25 mm below the cryostat, while most of the measured unenhanced signal came from the regions closer to the cryostat bottom, where the SQUID pick-up coil has greater sensitivity [19]. The DNP-enhanced NMR signal was found to decay as approximately a linear function of time, dropping by a factor of 3 in 20 min. Such quasi-linear decrease in signal intensity has been observed in ESR studies involving nitroxide free radicals [42]. It is due to a gradual reduction of nitroxides in biological systems (for example, via conversion to hydroxylamines), which results in the loss of paramagnetism [42]. Because such reduction depends on chemical environment, it has been used as a sensitive tool to study various biochemical processes [41]. Our experiment shows that ULF MRI with Overhauser DNP can be successfully used for imaging small animals and plants. It also suggests that 3D Fourier encoding is too time-consuming for this purpose, and imaging sequences with slice selection should be used instead.

*3.3 Overhauser-enhanced ULF NMR of carbon-13*

As mentioned in Sec. 1, ULF NMR with its broadband SQUID-based signal reception allows simultaneous measurement of NMR signals from $^1$H and $^{13}$C at microtesla-range magnetic fields. NMR of $^{13}$C, however, is considerably more challenging. For a substance that is labeled with $^{13}$C in one position within the molecule and has a typical concentration of 1 M in water solution, the $^{13}$C magnetization is ~1700 times lower than the magnetization of protons in water. Nevertheless, the DNP polarization enhancement makes it possible to perform NMR spectroscopy of $^{13}$C at microtesla fields.

The first NMR measurement of $^{13}$C at microtesla fields is demonstrated in Fig.8. The experiment was performed as described in Sec. 2.4. TEMPO concentration of 16 mM was found to provide the strongest $^{13}$C signal for 1 M concentration of sodium bicarbonate. The $^{13}$C NMR signal could only be observed with DNP enhancement. We estimated the unenhanced $^{13}$C signal indirectly using the unenhanced signal from the solvent water protons as a reference, and taking into account the difference in spin concentration, $N(^1\text{H})/N(^{13}\text{C})\sim 110$, the difference in magnetization (for the same concentration), $\gamma^2(^1\text{H})/\gamma^2(^{13}\text{C})\sim 16$, the difference in measured linewidth, $T_2^*(^{13}\text{C})/T_2^*(^1\text{H})\sim 17$, and the decrease in $^1$H signal during the field switching (Sec. 2.4). The polarization enhancement factor for $^{13}$C was thus estimated to be $E \approx -200$. For $B_p=5.75$ mT, this corresponds to the equivalent polarization field of 1.1 T. The longitudinal relaxation time for $^{13}$C in sodium bicarbonate with 16 mM TEMPO was measured to be $T_1=3.0\pm 0.1$ s at 96 µT and $T_1=3.5\pm 0.1$ s at 5.75 mT. The pre-polarization time was $t_p=5.0$ s. The described estimation of the unenhanced $^{13}$C signal from the unenhanced $^1$H signal corresponded to $t_p \gg T_1$, as for the water protons in the presence of 16 mM TEMPO. If the same condition held for $^{13}$C in the DNP experiment, the measured $^{13}$C NMR signal would likely be ~30% higher, leading to a larger estimated enhancement factor $E \approx -260$.

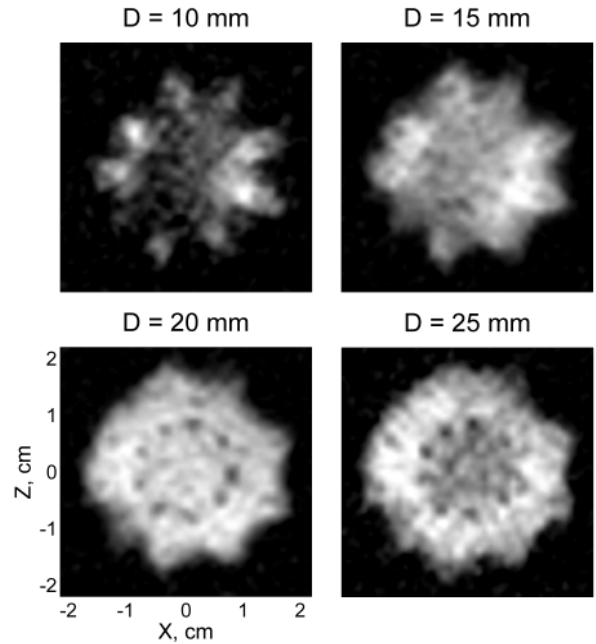

**Fig. 7.** DNP-enhanced 3D image of the moon cactus, injected with TEMPO solution. The image was acquired at 96 µT in a single scan with 1 mm × 1 mm in-plane resolution. $D$ is the depth of a given 5 mm thick layer under the bottom of the cryostat.

$^{13}$C NMR spectra of [1-$^{13}$C] sodium pyruvate and [1-$^{13}$C] L-alanine at the 96 µT field are exhibited in Fig. 9, together with the $^{13}$C sodium bicarbonate spectrum from Fig. 8. All the spectra are plotted with the same vertical scale. The quartet splitting for pyruvate at 4 mM TEMPO in Fig. 9 is about 1.4 Hz, which is close to the value of $^3J_{CH} \sim 1.3$ Hz reported in [47]. The first observation one can make is that the $J$-coupling spectrum of pyruvate changes drastically with the concentration of free radicals: the quartet at 4 mM TEMPO becomes the broadened singlet at 16 mM. This behavior can be attributed to the effect of spin exchange (see, e.g. [49]). As the free radical concentration increases, the $T_1$ relaxation time for $^1$H spins (as well as for the $^{13}$C spin) in the pyruvate molecule is reduced. This leads to a faster $^1$H spin flipping between its two states, so the effect of $J$-coupling between $^1$H and $^{13}$C is increasingly averaged

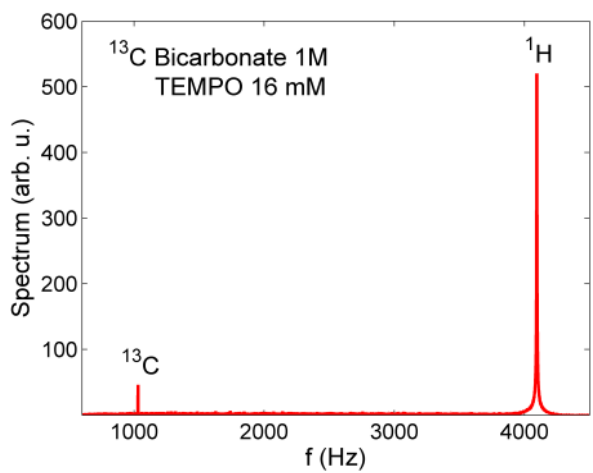

**Fig. 8.** Simultaneous measurement of proton and carbon-13 NMR signals at 96 µT field using Overhauser DNP.

out. As a result, the sharp multiplet in the limit of $1/T_1^H \ll 2\pi J_{CH}$ (slow spin exchange) becomes a sharp singlet in the $1/T_1^H \gg 2\pi J_{CH}$ limit (fast spin exchange). In the intermediate regime, the multiplet lines are broadened and their splitting is reduced, until the multiplet collapses into a broadened singlet, which becomes sharper as the spin exchange rate is increased [49]. Such "chemical exchange spin decoupling" has been observed and studied before [50−52].

Another result to be noted in Fig. 9 is that the $^{13}$C spectrum of L-alanine is a broadened singlet instead of the expected octet (Sec. 2.5) even at 4 mM TEMPO concentration. This means that the decoupling effect of TEMPO is stronger for alanine than for pyruvate, which can be explained as follows. Nitroxide free radicals are known to form hydrogen bonds with protons in −NH$_2$, −OH, and −COOH groups [53]. Such bonding causes faster $T_1$ relaxation for $^1$H and $^{13}$C in the case of alanine, leading to the stronger decoupling effect. Relaxation enhancement due to hydrogen bond formation has also been observed with paramagnetic agents ([54], p.226). This explanation for the appearance of the alanine spectrum in Fig. 9 is further supported by $T_1$ measurements for $^{13}$C. For pyruvate, $T_1$ values were determined to be 3.4±0.3 s (at 96 µT) and 5.0±0.3 s (at 5.7 mT) for 16 mM TEMPO concentration. These values increased to 6.2±0.5 s (96 µT) and 8.7±0.5 s (5.7 mT) for 4 mM TEMPO. In the case of alanine, the corresponding $T_1$ values were measured to be 3.9±0.4 s (96 µT) and 4.6±0.4 s (5.7 mT) at 4 mM TEMPO. Because the NMR signals from pyruvate and alanine were very weak (Fig. 9), these numbers are rather coarse estimates obtained by fitting an exponential function to only three data points as in [22]. Comparison of the $T_1$ relaxation times suggests that the 4 mM TEMPO concentration has approximately the same effect on $^{13}$C relaxation properties of alanine as the 16 mM of TEMPO – on the relaxation properties of pyruvate.

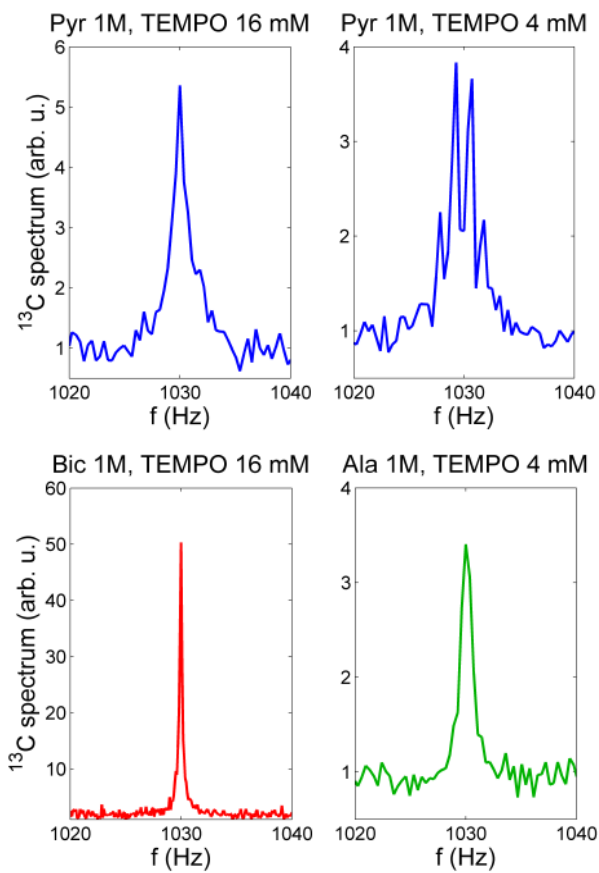

**Fig. 9.** Carbon-13 spectra of sodium bicarbonate, sodium pyruvate, and alanine, measured at 96 µT field using DNP. The effect of TEMPO concentration on *J*-coupling spectrum is clearly observed in the case of pyruvate. Similarly, the *J*-coupling spectrum of alanine, normally a multiplet, appears as a singlet in the figure (see text for details).

We also tried to carry out Overhauser-enhanced ULF NMR of [1-$^{13}$C] sodium lactate, but were unable to detect any $^{13}$C signal. This negative result can be attributed to the lower DNP enhancement [55], and, possibly, stronger broadening effects of TEMPO free radicals in the case of the lactate.

## 4. Discussion

Our results demonstrate that Overhauser DNP at mT-range magnetic fields can be naturally combined with ULF NMR/MRI in one system. This approach greatly improves SNR of ULF NMR/MRI by increasing nuclear spin polarization by two orders of magnitude without an increase in the pre-polarizing field strength.

Our results also suggest that one should exercise caution when evaluating DNP performance based on the enhancement factor *E*. In Sec. 3.1, polarization enhancement with $E = -95$ was observed for protons at $B_p = 3.52$ mT ($T_{25\pi}$ transition). If the same measurements were carried out at 2.15 mT ($T_{16\pi}$ transition) and 5.77 mT ($T_{34\pi}$ transition), and the resulting non-equilibrium polarizations were compared to the thermal polarizations at those fields, the enhancement factors would likely be around −150 and −60, respectively. However, the absolute polarization levels in all three cases would be similar (Fig. 5A), corresponding to the thermal polarization at ~0.3 T magnetic field. Values of the factor $E$ may be more confusing when the DNP-enhanced polarization at one magnetic field is compared to the thermal equilibrium polarization at a different field strength. In [44], for example, the authors performed Overhauser DNP at 2.7 mT field ($T_{16\pi}$ transition at 131.5 MHz), compared the resulting NMR signal to the unenhanced signal at the Earth magnetic field of ~54 µT, and reported $E = -3100$. If the same 2.7 mT magnetic field were used as a reference, the enhancement factor value would be $E = -62$. Therefore, DNP performance is more accurately characterized by the strength of the equivalent magnetic field that would produce the same thermal equilibrium polarization as the measured DNP-enhanced polarization. In our experiments, this equivalent field is 0.33 T for protons, and 1.1 T for $^{13}$C.

Microtesla MRI with Overhauser enhancement and SQUID signal detection, described in Sec. 3.2, has an important advantage over conventional Overhauser-enhanced MRI, which relies on Faraday signal detection. Because sensitivity of Faraday detection scales linearly with Larmor frequency, a stronger detection field is applied in conventional Overhauser-enhanced MRI after the Overhauser DNP at a low magnetic field. In [31], for instance, the DNP was performed at 60 µT and MR images were acquired at 6.8 mT. In the most advanced system, designed for in vivo imaging of free radicals [30], Overhauser DNP at each imaging step is carried out at 5 mT, and the magnetic field is then ramped up to 0.45 T for MRI signal detection. In our approach, imaging is performed at the microtesla-range measurement field $B_m$ (Fig. 3A) using highly sensitive SQUIDs, so a stronger detection field is not needed. This allows development of more efficient systems for Overhauser-enhanced MRI, utilizing magnetic fields no higher than a few mT.

The NMR experiments, described in Sec. 3.3, show that the combination of Overhauser DNP with SQUID signal detection makes it possible to perform NMR of $^{13}$C at microtesla-range magnetic fields. Very few studies have been reported in which $^{13}$C NMR signals were measured below 1 T. In [56], Overhauser DNP and NMR of $^{13}$C were carried out at 7.4 mT field, but that work used $^{13}$C containing solvents rather than water solutions of $^{13}$C chemicals. More recently, Overhauser-enhanced NMR measurements of $^{13}$C labeled materials in aqueous solutions were reported at 0.35 T magnetic field [57]. The NMR spectroscopy of $^{13}$C at microtesla fields, demonstrated here for the first time, can be used to measure $J$-coupling constants for $^{13}$C spins in various substances with very high (mHz range) spectral resolution. Moreover, because SQUID sensors with untuned input circuits enable broadband signal reception, NMR spectra of various nuclear spins, including $^{13}$C, $^{15}$N, $^{31}$P, $^{19}$F, and $^1$H, can be measured simultaneously with the same detection sensitivity. Magnetic relaxometry of these spins can provide important information about molecular dynamics. Clearly, both $J$-coupling and magnetic relaxation properties of $^{13}$C are strongly affected, if free radicals are present in the solution (Fig. 9). This limitation can be overcome in continuous flow systems, in which a solution of $^{13}$C labeled material under study is pumped through a special porous media (gel) chemically labeled with stable radicals [58]. This would allow the studied solution to be separated from the radicals after the DNP stage and before the NMR measurement.

The potential of microtesla MRI in combination with DNP is not limited to the case of Overhauser DNP. The most promising integration approach, in our opinion, is combination of SQUID-based microtesla MRI (without pre-polarization) with the $^{13}$C hyperpolarization method [34]. Because the hyperpolarization is performed outside an MRI scanner using a separate NMR-style hyperpolarizer, such as HyperSense® from Oxford Instruments (www.oxford-instruments.com), high magnetic fields of conventional MRI systems offer little advantage in terms of achievable $^{13}$C polarization. The same applies to the pre-polarizing field $B_p$ of conventional ULF MRI. It has been suggested that "the use of hyperpolarized substances should make it possible to perform imaging at magnetic field, lower than the one used in clinical routine of today, and still generate images with high SNR" [37]. We propose to perform imaging of an injected hyperpolarized material labeled with $^{13}$C using a novel type of MRI scanner (Fig. 10), that employs only microtesla-range magnetic fields.

The scanner in Fig. 10 has several features that could make it an attractive option for imaging hyperpolarized $^{13}$C (or $^{15}$N). First, it can be much more open, portable, and inexpensive than conventional MRI machines. Moreover, because a DNP hyperpolarizer can produce a large (~100 ml) amount of the hyperpolarized solution [59], several microtesla MRI scanners can be operated simultaneously after a single hyperpolarizer's run. Second, because this scanner only employs magnetic fields below ~1 mT, SQUID sensors of any type can be successfully used for MRI signal detection without the risk of trapping magnetic flux. This greatly simplifies the SQUID system design, because the special cryoswitches, employed in ULF MRI to protect the SQUIDs from trapping flux during the strong $B_p$ pulses [19], are no longer needed. Third, because hyperpolarized $^{13}$C provides a strong MRI signal for a relatively short time, parallel imaging with a large SQUID array can be efficiently used to accelerate image acquisition [20]. Modern magnetoencephalography (MEG) systems include as many as ~300 SQUID channels, and their technology is well developed. The image acquisition can be accelerated even further if the MEG-style source localization by a large sensor array is used in place of the conventional MRI encoding. Even though chemical shift information is not available in the low-field limit, metabolites of an injected hyperpolarized substance can, in principle, be distinguished by means of $J$-coupling spectroscopy and/or magnetic relaxometry of $^{13}C$ at microtesla fields. Our experimental results in Sec. 3.3, together with the published low-resolution spectra for [1-$^{13}$C] alanine and lactate [48] suggest that this might indeed be possible. Therefore, the proposed ultimate low-field MRI scanner (Fig. 10) can potentially be used not only for angiography and perfusion mapping, but also for real-time metabolic imaging with hyperpolarized $^{13}C$.

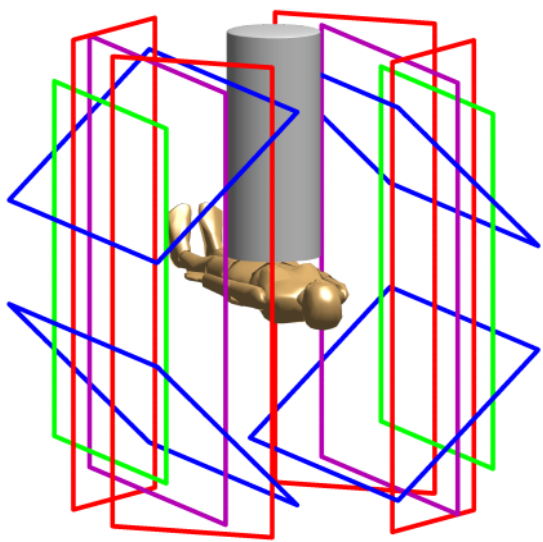

**Fig. 10.** Conceptual drawing of an ultimate low-field MRI scanner for imaging hyperpolarized substances. It uses only microtesla-range magnetic fields, and is open, portable, and inexpensive. Color coding for the coil sets is the same as in Fig. 2A.

## 5. Conclusion

Based on our experimental results, presented in this work, we conclude that combination of DNP with SQUID detection greatly enhances SNR performance of low-field NMR/MRI. This approach can be used to develop new types of research instruments for in vivo imaging and in vitro materials studies. They include: i) an efficient system for Overhauser-enhanced MRI, with Overhauser DNP at a mT-range and image acquisition at a µT-range magnetic field; ii) a continuous flow system for Overhauser-enhanced NMR of $^{13}C$, suitable for both $J$-coupling spectroscopy at µT fields and $T_1$ relaxometry in the µT − mT field range; iii) an open MRI scanner for imaging DNP hyperpolarized $^{13}C$ using only µT-range magnetic fields. Such instruments could have important applications, particularly in studies of metabolism involving free radicals or endogenous substances labeled with $^{13}C$.

## Acknowledgments

We would like to thank Joe Bradley, John Galbraith, Kevin Henderson, Tom Lopez, Mark Peters, Dale Tupa, and Algis Urbaitis for their technical help. We also gratefully acknowledge the support of the U.S. National Institutes of Health Grant R01-EB006456 and of the U.S. Department of Energy Office of Biological and Environmental Research Grant KP1503020.

## References

[1] D.I. Hoult, R.E. Richards, The signal-to-noise ratio of the nuclear magnetic resonance experiment, J. Magn. Reson. 24 (1976) 71-85.
[2] M. Packard, R. Varian, Free nuclear induction in the Earth's magnetic field, Phys. Rev. 93 (1954) 941.
[3] A. Mohorič, J. Stepišnik, NMR in the Earth's magnetic field, Progr. Nucl. Magn. Reson. Spectrosc. 54 (2009) 166-182.
[4] S. Appelt, H. Kühn, F.W. Häsing, B. Blümich, Chemical analysis by ultrahigh-resolution nuclear magnetic resonance in the Earth's magnetic field, Nature Phys. 2 (2006) 105-109.
[5] J.N. Robinson, A. Coy, R. Dykstra, C.D. Eccles, M.W. Hunter, P.T. Callahan, Two-dimensional NMR spectroscopy in Earth's magnetic field, J. Magn. Reson. 182 (2006) 343-347.
[6] R. McDermott, A.H. Trabesinger, M. Mück, E.L. Hahn, A. Pines, J. Clarke, Liquid-state NMR and scalar couplings in microtesla magnetic fields, Science 295 (2002) 2247-2249.
[7] J. Clarke, M. Hatridge, M. Mößle, SQUID-detected magnetic resonance imaging in microtesla fields, Annu. Rev. Biomed. Eng. 9 (2007) 389-413.
[8] J. Clarke and A.I. Braginski (Eds.), The SQUID Handbook, Wiley-VCH, Weinheim, 2004.
[9] Ya.S. Greenberg, Application of superconducting quantum interference devices to nuclear magnetic resonance, Rev. Mod. Phys. 70 (1998) 175-222.
[10] A.H. Trabesinger, R. McDermott, S.K. Lee, M. Mück, J. Clarke, A. Pines, SQUID-detected liquid state NMR in microtesla fields, J. Chem. Phys. A 108 (2004) 957-963.
[11] A.N. Matlachov, P.L. Volegov, M.A. Espy, J.S. George, R.H. Kraus Jr., SQUID detected NMR in microtesla magnetic fields, J. Magn. Reson. 170 (2004) 1-7.
[12] P.L. Volegov, A.N. Matlashov, R.H. Kraus Jr., Ultra-low field NMR measurements of liquids and gases with short relaxation times, J. Magn. Reson. 183 (2006) 134-141.
[13] R.H. Kraus Jr, P. Volegov, A. Matlachov, M. Espy, Toward direct neural current imaging by resonant mechanisms at ultra-low field, NeuroImage 39 (2008) 310-317.
[14] I. Savukov, A. Matlashov, P. Volegov, M. Espy, M. Cooper, Detection of $^3$He spins with ultra-low field nuclear magnetic resonance employing SQUIDs for application to a neutron electric dipole moment experiment, J. Magn. Reson. 195 (2008) 129-133.
[15] P.E. Magnelind, A.N. Matlashov, P.L. Volegov, M.A. Espy, Ultra-low field NMR of UF$_6$ for $^{235}$U detection and characterization, IEEE Trans. Appl. Supercond. 19 (2009) 816-818.

[16] M. Burghoff, H.H. Albrecht, S. Hartwig, I. Hilschenz, R. Körber, T.S. Thömmes, J.J. Scheer, J. Voigt, L. Trahms, SQUID system for MEG and low field magnetic resonance, Metrol. Meas. Syst. 16 (2009) 371-375.

[17] L. Qiu, Y. Zhang, H.J. Krause, A.I. Braginski, A. Offenhäusser, Low-field NMR measurement procedure when SQUID detection is used, J. Magn. Reson. 196 (2009) 101-104.

[18] R. McDermott, S.K. Lee, B. ten Haken, A.H. Trabesinger, A. Pines, J. Clarke, Microtesla MRI with a superconducting quantum interference device, Proc. Nat. Acad. Sci. USA 101 (2004) 7857-7861.

[19] V.S. Zotev, A.N. Matlashov, P.L. Volegov, A.V. Urbaitis, M.A. Espy, R.H. Kraus Jr., SQUID-based instrumentation for ultralow-field MRI, Supercond. Sci. Technol. 20 (2007) S367-S373.

[20] V.S. Zotev, P.L. Volegov, A.N. Matlashov, M.A. Espy, J.C. Mosher, R.H. Kraus Jr., Parallel MRI at microtesla fields, J. Magn. Reson. 192 (2008) 197-208.

[21] V.S. Zotev, A.N. Matlashov, P.L. Volegov, I.M. Savukov, M.A. Espy, J.C. Mosher, J.J. Gomez, R.H.Kraus, Jr., Microtesla MRI of the human brain combined with MEG, J. Magn. Reson. 194 (2008), 115-120.

[22] V.S. Zotev, A.N. Matlashov, I.M. Savukov, T. Owens, P.L. Volegov, J.J. Gomez, M.A. Espy, SQUID-based microtesla MRI for in vivo relaxometry of the human brain, IEEE Trans. Appl. Supercond. 19 (2009) 823-826.

[23] I.M. Savukov, V.S. Zotev, P.L. Volegov, M.A. Espy, A.N. Matlashov, J.J. Gomez, R.H. Kraus Jr., MRI with an atomic magnetometer suitable for practical imaging applications, J. Magn. Reson. 199 (2009) 188-191.

[24] M. Espy et al., Ultra-low-field MRI for the detection of liquid explosives, Supercond. Sci. Technol. 23 (2010) 034023 (1-8).

[25] H.C. Yang, H.E. Horng, S.Y. Yang, S.H. Liao, Advances in biomagnetic research using high-$T_c$ superconducting quantum interference devices, Supercond. Sci. Technol. 22 (2009) 093001 (1-13).

[26] A.W. Overhauser, Polarization of nuclei in metals, Phys. Rev. 92 (1953) 411-415.

[27] K.H. Haussser, D. Stehlik, Dynamic nuclear polarization in liquids, Adv. Magn. Reson. 3 (1968) 79-139.

[28] W. Müller-Warmuth, K. Meise-Gresch, Molecular motions and interactions as studied by dynamic nuclear polarization (DNP) in free radical solutions, Adv. Magn. Reson. 11 (1983) 1-45.

[29] D.J. Lurie, D.M. Bussel, L.H. Bell, J.R. Mallard, Proton-electron double magnetic resonance imaging of free radical solutions, J. Magn. Reson. 76 (1988) 366-370.

[30] D.J. Lurie, G.R. Davies, M.A. Foster, J.M.S. Hutchison, Field-cycled PEDRI imaging of free radicals with detection at 450 mT, Magn. Reson. Imaging 23 (2005) 175-181.

[31] G. Planinšič, T. Guibertau, D. Grucker, Dynamic nuclear polarization imaging in very low magnetic fields, J. Magn. Reson. B 110 (1996) 205-209.

[32] M.C. Krishna et al., Overhauser enhanced magnetic resonance imaging for tumor oximetry: coregistration of tumor anatomy and tissue oxygen concentration, Proc. Natl. Acad. Sci. USA 99 (2002) 2216-2221.

[33] S. Matsumoto, K. Yamada, H. Hirata, K. Yasukawa, F. Hyodo, K. Ichikawa, H. Utsumi, Advantageous application of a surface coil to EPR irradiation in Overhauser-enhanced MRI, Magn. Reson. Med. 57 (2007) 806-811.

[34] J.H. Ardenkjaer-Larsen, B. Fridlund, A. Gram, G. Hansson, L. Hansson, M.L. Lerche, R. Servin, M. Thaning, K. Golman, Increase in signal-to-noise ratio of >10,000 times in liquid-state NMR, Proc. Natl. Acad. Sci. USA 100, (2003) 10158-10163.

[35] A. Comment, B. van den Brandt, K. Uffmann, F. Kurdzesau, S. Jannin, J.A. Konter, P. Hautle, W.T. Wenckebach, R. Gruetter, J.J. van der Klink, Design and performance of a DNP prepolarizer coupled to a rodent MRI scanner, Concepts Magn. Reson. B 31 (2007) 255-269.

[36] S. Mansson, E. Johansson, P. Magnusson, C.M. Chai, G. Hansson, J.S. Petersson, F. Stahlberg, K. Golman, $^{13}$C imaging – a new diagnostic platform, Eur. Radiol. 16 (2006) 57-67.

[37] K. Golman, J.S. Petersson, Metabolic imaging and other applications of hyperpolarized $^{13}$C, Acad. Radiol. 13 (2006) 932-942.

[38] F.A. Gallagher, M.I. Kettunen, K.M. Brindle, Biomedical applications of hyperpolarized $^{13}$C magnetic resonance imaging, Progr. Nucl. Magn. Reson. Spectrosc. 55 (2009) 285-295.

[39] K.P. Pruessmann, Medical imaging: less is more, Nature 455 (2008) 43-44.

[40] I. Nicholson, D.J. Lurie, F.J.L. Robb, The application of proton-electron double-resonance imaging techniques to proton mobility studies, J. Magn. Reson. B 104 (1994), 250-255.

[41] G.I. Likhtenshtein, J. Yamauchi, S. Nakatsuji, A.I. Smirnov, and R. Tamura, Nitroxides – Applications in Chemistry, Biomedicine, and Materials Science, Wiley-VCH, Weinheim, 2008.

[42] N. Kocherginsky, H.M. Swartz, Nitroxide Spin Labels – Reactions in Biology and Chemistry, CRC Press, Boca Raton, 1995.

[43] T. Guibertau, D. Grucker, EPR spectroscopy by dynamic nuclear polarization in low magnetic field, J. Magn. Reson. B 110 (1996), 47-54.

[44] M.E. Halse, P.T. Callaghan, A dynamic nuclear polarization strategy for multi-dimensional Earth's field NMR spectroscopy, J. Magn. Reson. 195 (2008) 162-168.

[45] B.D. Armstrong, S. Han, A new model for Overhauser enhanced nuclear magnetic resonance using nitroxide radicals, J. Chem. Phys. 127 (2007) 104508 (1-10).

[46] H. Hirata, T. Walczak, H.M. Swartz, Electronically tunable surface-coil-type resonator for L-band EPR spectroscopy, J. Magn. Reson. 142 (2000) 159-167.

[47] S.G. Lloyd, H. Zeng, P. Wang, J.C. Chatham, Lactate isotopomer analysis by $^1$H NMR spectroscopy: consideration of long-range nuclear spin-spin interactions, Magn. Reson. Med. 51 (2004) 1279-1282.

[48] Y.S. Levin, D. Mayer, Y.F. Yen, R.E. Hurd, D.M. Spielman, Optimization of fast spiral chemical shift imaging using least squares reconstruction: application for hyperpolarized $^{13}$C metabolic imaging, Magn. Reson. Med. 58 (2007) 245-252.

[49] E.D. Becker, High Resolution NMR – Theory and Chemical Applications, Academic Press, San Diego, 1999.

[50] L.S. Frankel, NMR investigation of relaxation and "chemical-exchange spin decoupling" in dilute solutions of $Co^{2+}$ with various "P-O" containing solvents, J. Chem. Phys. 50 (1969) 943-950.

[51] U. Sequin, A.I. Scott, Chemical suppression of long range $^{13}$C-$^1$H coupling in $^{13}$C NMR spectra, J.C.S. Chem. Comm. (1974) 1041-1042.

[52] I. Morishima, T. Inubushi, S. Uemura, H. Miyoshi, Nitroxide radical as a nuclear spin decoupling reagent. Application to carbon-13 nuclear magnetic resonance studies of organothallium compounds, J. Am. Chem. Soc. 100 (1978) 354-356.

[53] J.L. Russ, J. Gu, K.H. Tsai, T. Glass, J.C. Duchamp, H.C. Dorn, Nitroxide/substrate weak hydrogen bonding: attitude and dynamics of collisions in solution, J. Am. Chem. Soc. 129 (2007) 7018-7027.

[54] F.W. Wehrli, A.P. Marchand, and S. Wehrli, Interpretation of Carbon-13 NMR Spectra, John Wiley and Sons, 1988.

[55] D.K. Deelchand, I. Iltis, M. Marjanska, C. Nelson, K. Ugurbil, P.G. Henry, Localized detection of hyperpolarized [1-$^{13}$C] pyruvate and its metabolic products in rat brain, Proc. Intl. Soc. Magn. Reson. Med., Toronto, 2008, p.3196.

[56] R.D. Bates, B.E. Wagner, E.H. Poindexter, Dynamic polarization of carbon-13 nuclei in free radical solutions, Chem. Phys. Lett. 17 (1972) 328-331.
[57] M.D. Lingwood, S. Han, Dynamic nuclear polarization of $^{13}$C in aqueous solutions under ambient conditions, J. Magn. Reson. 201 (2009) 137-145.
[58] M.D. Lingwood, T.A. Siaw, N. Sailasuta, B.D. Ross, P. Bhattacharya, S. Han, Continuous flow Overhauser dynamic nuclear polarization of water in the fringe field of a clinical magnetic resonance imaging system for authentic image contrast, J. Magn. Reson. 205 (2010) 247-254.
[59] A. Comment, J. Rentsch, F. Kurdzesau, S. Jannin, K. Uffmann, R.B. van Heeswijk, P. Hautle, J.A. Konter, B. van den Brandt, J.J. van der Klink, Producing over 100 ml of highly concentrated hyperpolarized solution by means of dissolution DNP, J. Magn. Reson. 194 (2008) 152-155.